\documentstyle[aps,prl,multicol,epsfig]{revtex}
\draft
\begin{document}
\title{Magnetic field--induced Kondo effects in Coulomb blockade
systems} 
\author{M. Pustilnik$^a$, L. I. Glazman$^a$,
D. H. Cobden$^b$, and L. P. Kouwenhoven$^c$ 
\vspace{3mm} } 
\address{
$^a$ Theoretical Physics Institute, University of Minnesota, 116
Church St. SE, Minneapolis, MN 55455 \\ 
$^b$ Department of Physics,
University of Warwick, Coventry, CV4 7AL, UK \\ 
$^c$ Department of
Applied Physics and ERATO Mesoscopic Correlation Project,\\ 
Delft University of Technology, P. O. Box 5046, 2600 GA Delft, the
Netherlands \vspace{2.5mm}} 
\date{\today\vspace{2.5mm}} 
\maketitle
\begin{abstract}
We review the peculiarities of transport through a quantum dot caused
by the spin transition in its ground state. Such transitions can be
induced by a magnetic field. Tunneling of electrons between the dot
and leads mixes the states belonging to the ground state manifold of
the dot. Unlike the conventional Kondo effect, this mixing, which
occurs only at the singlet-triplet transition point, involves both the
orbital and spin degrees of freedom of the electrons. We present
theoretical and experimental results that demonstrate the enhancement
of the conductance through the dot at the transition point.
\end{abstract}


\vspace{2mm}
\begin{multicols}{2}
\section{Introduction} 
Quantum dot devices provide a well--controlled object for studying
quantum many-body physics. In many respects, such a device resembles
an atom imbedded into a Fermi sea of itinerant electrons. These
electrons are provided by the leads attached to the dot. The orbital
mixing in the case of quantum dot corresponds to the electron
tunneling through the junctions connecting the dot with leads. Voltage
$V_g$ applied to a gate -- an electrode coupled to the dot
capacitively -- allows one to control the number of electons $N$ on
the dot. Almost at any gate voltage an electron must have a finite
energy in order to overcome the on-dot Coulomb repulsion and tunnel
into the dot. Therefore, the conductance of the device is suppressed
at low temperatures (Coulomb blockade phenomenon\cite{Leo}). The
exceptions are the points of charge degeneracy. At these points, two
charge states of the dot have the same energy, and an electron can hop
on and off the dot without paying an energy penalty. This results in a
periodic peak structure in the dependence of the conductance $G$ on
$V_g$. Away from the peaks, in the Coulomb blockade valleys, the
charge fluctuations are negligible, and the number of electrons $N$ is
integer.

Every time $N$ is tuned to an odd integer, the dot must carry a
half-integer spin. In the simplest case, the spin is $S=1/2$, and is
due to a single electron residing on the last occupied discrete level
of the dot. Thus, the quantum dot behaves as $S=1/2$ magnetic impurity
imbedded into a tunneling barrier between two massive conductors. It
is known\cite{reviews} since mid-60's that the presence of such
impurities leads to zero-bias anomalies in tunneling
conductance\cite{classics}, which are adequately
explained\cite{Appelbaum} in the context of the Kondo
effect\cite{kondo}. The advantage of the new experiments\cite{exp1} is
in full control over the ``magnetic impurity'' responsible for the
effect. For example, by varying the gate voltage, $N$ can be
changed. Kondo effect results in the increased low--temperature
conductance only in the odd--$N$ valleys. The even--$N$ valleys
nominally correspond to the $S=0$ spin state (non-magnetic impurity),
and the conductance decreases with lowering the temperature.

Unlike the real atoms, the energy separation between the discrete
states in a quantum dot is fairly small. Therefore, the $S=0$ state of
a dot with even number of electrons is much less robust than the
corresponding ground state of a real atom. Application of a magnetic
field in a few--Tesla range may result in a transition to a
higher-spin state.  In such a transition, one of the electrons
residing on the last doubly--occupied level is promoted to the next
(empty) orbital state. The increase in the orbital energy accompanying
the transition is compensated by the decrease of Zeeman and exchange
energies. At the transition point, the ground state of the dot is
degenerate. Electron tunneling between the dot and leads results in
mixing of the components of the ground state. Remarkably, the mixing
involves spin as well as orbital degrees of freedom. In this paper we
demonstrate that the mixing yields an enhancement of the
low--temperature conductance through the dot. This enhancement can be
viewed as the magnetic--field--induced Kondo effect.

We present the model and theory of electron transport in the
conditions of the field-induced Kondo effect in Section~\ref{MODEL}. The
experimental manifestations of the transition observed on GaAs
vertical quantum dots and carbon nanotubes are described in Section~\ref{EXP}.

\section{The model\label{MODEL}}
\begin{figure}
\centerline{\epsfxsize=5.5cm
\epsfbox{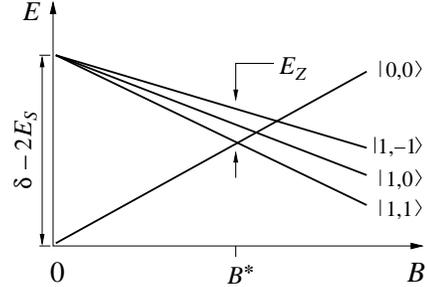}\vspace{1.5mm}}
\caption{Typical picture of the singlet-triplet transition in the ground state of 
a quantum dot.}
\label{crossings}
\end{figure}

We will be considering a confined electron system which does not have
special symmetries, and therefore the single-particle levels in it are
non-degenerate. In addition, we assume the electron-electron
interaction to be relatively weak (the gas parameter $r_s\lesssim
1$). Therefore, discussing the ground state, we concentrate on the
transitions which involve only the lowest-spin states. In the case of
even number of electrons, these are states with $S=0$ or $S=1$. At a
sufficiently large level spacing
$\delta\equiv\epsilon_{+1}-\epsilon_{-1}$ between the last occupied
($-1$) and the first empty orbital level ($+1$), the ground state is a
singlet at $B=0$. Finite magnetic field affects the orbital energies;
if it reduces the difference between the energies of the said orbital
levels, a transition to a state with $S=1$ may occur, see 
Fig.~\ref{crossings}. Such a transition involves rearrangement of two 
electrons between the levels $n=\pm 1$. Out of the six states involved, 
three belong to a triplet $S=1$, and three others are singlets ($S=0$). 
The degeneracy of the triplet states is removed only by Zeeman energy.  
The singlet states, in general, are not degenerate with each other. To 
describe the transition between a singlet and the triplet in the ground state, 
it is sufficient to consider the following Hamiltonian:
\begin{equation}
H_{\rm dot}=\sum_{ns}\epsilon _{n}d_{ns}^{\dagger }d_{ns}
-E_{S}{\bf S}^2 - E_Z S^z 
+E_{C}\left( N-{\cal N}\right) ^{2}.  
\label{Hdot}
\end{equation}
Here, $N=\sum_{s,n}d_{ns}^{\dagger }d_{ns}$ is the total number of
electrons occupying the levels $n=\pm 1$, operator 
${\bf S} = \sum_{nss'}d_{ns}^{\dagger }
\left( \bbox{\sigma }_{ss^{\prime }}/2\right) d_{ns'}$ 
is the corresponding total spin ($\bbox{\sigma}$ are the Pauli matrices), 
and the parameters $E_S$, $E_Z=g\mu_B B$, and $E_{C}$ 
are the exchange, Zeeman, and charging energies respectively \cite{exchange}.  
We restrict our attention to the very middle of a Coulomb blockade valley 
with an even number of electrons in the dot (that is modelled by setting 
the dimensionless gate voltage ${\cal N}$ to ${\cal N}=2$).  We assume 
that the level spacing $\delta$ is tunable, {\it e.g.}, by means of a magnetic 
field $B$: $\delta=\delta(B)$, and that $\delta(0) > 2E_S$ (which ensures that
the dot is non-magnetic for $B=0$).  

The lowest--energy singlet state and the three components of the competing 
triplet state can be labeled as $\left| S,S^{z}\right\rangle $ in terms of the
total spin $S$ and its $z$--projection $S_z$,
\begin{eqnarray}
&&|1,1\rangle =d_{+1\uparrow }^{\dagger }d_{-1\uparrow }^{\dagger}|0\rangle, 
\nonumber \\
&&|1,-1\rangle =d_{+1\downarrow}^{\dagger }d_{-1\downarrow }^{\dagger }
|0\rangle ,  
\label{Basis} \\
&&|1,0\rangle =\frac{1}{\sqrt{2}}\left( d_{+1\uparrow }^{\dagger
}d_{-1\downarrow }^{\dagger }+d_{+1\downarrow }^{\dagger }d_{-1\uparrow
}^{\dagger }\right) | 0\rangle ,  
\nonumber \\
&&|0,0\rangle =d_{-1\uparrow }^{\dagger }d_{-1\downarrow}^{\dagger }
|0\rangle,   
\nonumber
\end{eqnarray}
where $|0\rangle $ is the state with the two levels empty.  According
to (\ref{Hdot}), the energies of these states satisfy
\begin{equation}
E_{|S,S^z\rangle} - E_{|0,0\rangle} 
= K_0 S - E_Z S^z, 
\label{Hd}
\end{equation}
where $K_0 = \delta-2E_S$.  Since $\delta > 2E_S$, the ground state of
the dot at $B=0$ is a singlet $|0,0\rangle$. Finite field shifts the
singlet and triplet states due to the orbital effect, and also leads
to Zeeman splitting of the components of the triplet. As $B$ is
varied, the level crossings occur (see Fig.~\ref{crossings}).  The
first such crossing takes place at $B=B^\ast$, satisfying the equation
\begin{equation}
\delta(B^\ast)- E_Z(B^\ast)=2E_S.
\label{Bast}
\end{equation}
At this point, the two states, $|0,0\rangle$ and $|1,1\rangle$, form a
doubly degenerate ground state, see Fig.~\ref{crossings}.

If leads are attached to the dot, the dot-lead tunneling results in
the hybridization of the degenerate (singlet and triplet) states. The
characteristic energy scale $T_0$ associated with the hybridization
can be in different relations with the Zeeman splitting at field
$B=B^*$. 

If $E_Z(B^*)\ll T_0$, then the Zeeman splitting between the triplet
states can be neglected, and at the $B=B^\ast$ point all {\it four}
states (\ref{Basis}) can be considered as degenerate. Theory for this
case is presented below in Section~\ref{TRANSITION}. This limit
adequately describes a quantum dot formed in a two-dimensional
electron gas (2DEG) at the GaAs-AlGaAs interface, subject to a
magnetic field, see Section~\ref{DOTS}. Energy $E_Z$ can be
neglected due to the smallness of the electron $g$-factor in GaAs.

Alternatively, the orbital effect of the magnetic field
($B$-dependence of $\delta$) may be very weak due to the reduced
dimensionality of the system, while the $g$-factor is not suppressed,
yielding an appreciable Zeeman effect even in a magnetic field of a
moderate strength. This limit of the theory, see Section~\ref{ZEEMAN},
corresponds to single-wall carbon nanotubes, which have very small widths of
about $1.4~ nm$ and $g = 2.0$. Measurements with carbon 
nanotubes are presented in Section~\ref{nanotubes}. 

In order to study the transport problem, we need to introduce into the
model the Hamiltonian of the leads and a term that describes the
tunneling. We choose them in the following form: 
\begin{eqnarray}
H_{l}&=&\sum_{\alpha nks}\xi _{k}c_{\alpha nks}^{\dagger }c_{\alpha nks},
\label{leads} \\
H_{T}&=&\sum_{\alpha nn'ks}t_{\alpha nn'}c_{\alpha nks}^{\dagger
  }d_{n's}+{\rm H.c.}
\label{HT} 
\end{eqnarray}
Here $\alpha =R,L$ for the
right/left lead, and $n=\pm 1$ for the two orbitals participating in the
singlet-triplet transition; $k$ labels states of the continuum spectrum in
the leads, and $s$ is the spin index. In writing (\ref{leads})-(\ref{HT}),
we had in mind the vertical dot device, where the potential creating
lateral confinement of electrons most probably does not vary much over the
thickness of the dot\cite{Tarucha}. Therefore we have assumed that the 
electron orbital motion perpendicular to the axis of the device can be
characterised by the same quantum number $n$ inside the dot and in the
leads. Presence of two orbital channels $n=\pm 1$ is important for the 
description of the Kondo effect at the singlet-triplet transition, that is, when 
the orbital effect of the magnetic field dominates. In the opposite case of 
large Zeeman splitting, the problem is reduced straightforwadly to the 
single-channel one, as we will see in Section~\ref{ZEEMAN} below.

\subsection{Effective Hamiltonian}
We will demonstrate the derivation of the effective low-energy Hamiltonian 
under the simplifying assumption \cite{EN},\cite{PG}
\begin{equation}
t_{\alpha nn'} = t_{\alpha}\delta_{nn'}.
\label{tunneling}
\end{equation}
This assumption, on one hand, greatly simplifies the calculations,
and, on the other hand, is still general enough to capture the most
important physical properties \cite{future}.

It is convenient to begin the derivation by performing a
rotation\cite{AM} in the R-L space
\begin{equation}
\left( 
\begin{array}{c}
\psi _{nks} \\ 
\phi _{nks}
\end{array}
\right) = 
\frac{1}{\sqrt {t_L^2 + t_R^2}}
\left( 
\begin{array}{cc}
t_{R} & t_{L} \\ 
-t_{L} & t_{R}
\end{array}
\right)
\left( 
\begin{array}{c}
c_{Rnks} \\ 
c_{Lnks}
\end{array}
\right),
\label{rotation}
\end{equation}
after which the $\phi$ field decouples: 
\begin{equation}
H_{T}=\sqrt {t_L^2 + t_R^2}\sum_{nks}\psi_{nks}^{\dagger}d_{ns}
+{\rm H.c.} 
\end{equation}
The differential conductance at zero bias $G$ can be related, using
Eq.~(\ref{rotation}), to the amplitudes of scattering
$A_{ns\rightarrow n's'}$ of the $\psi$--particles
\begin{equation}
G = \lim_{V\rightarrow 0}dI/dV
= \frac{e^2}{h} \left (\frac{2t_L t_R}{t_L^2 +  t_R^2}\right )^2
\sum _{nn'ss'} |A_{ns\rightarrow n's'}|^2.
\label{formula}
\end{equation}

The next step is to integrate out the virtual transitions to the
states with ${\cal N}\pm 1$ electrons by means of the Schrieffer-Wolff
transformation or, equivalently, by the Brillouin--Wigner perturbation 
theory. This procedure results in the effective low-energy Hamiltonian in 
which the transitions between the states (\ref{Basis}) are described by of 
the operators
\[
{\bf S}_{nn'}={\cal P}\sum_{ss'}d_{ns}^{\dagger }
\frac{\bbox{\sigma }_{ss'}}{2}
d_{n's'}{\cal P},
\]
where ${\cal P}= \sum_{S,S^z}|S,S^z\rangle\langle S,S^z|$ is the 
projection operator onto the system of states (\ref{Basis}). The operators 
${\bf S}_{nn'}$ may be conveniently written in terms of two fictitious 
$1/2$-spins ${\bf S}_{1,2}$. The idea of mapping comes from the 
one-to-one correspondence between the set of states (\ref{Basis}) 
and the states of  a two-spin system:
\begin{eqnarray*}
&&
|1,1\rangle \Longleftrightarrow |\uparrow_1 \uparrow_2 \rangle ,
\quad
|1,-1\rangle\Longleftrightarrow |\downarrow_1 \downarrow_2 \rangle ,
\\
&&
|1,0\rangle \Longleftrightarrow \frac{1}{\sqrt{2}}
\left(
|\uparrow_1 \downarrow_2 \rangle + |\downarrow_1 \uparrow_2 \rangle 
\right),
\\
&&
|0,0\rangle \Longleftrightarrow \frac{1}{\sqrt{2}}
\left(
|\uparrow_1 \downarrow_2 \rangle - |\downarrow_1 \uparrow_2 \rangle 
\right).
\end{eqnarray*}
We found the following relations:
\begin{eqnarray}
{\bf S}_{nn} &=&\frac{1}{2}\left( {\bf S}_{1}+{\bf S}_{2}\right) 
=\frac{1}{2}{\bf S}_{+},  
\nonumber \\
\sum_{n}{\bf S}_{-n,n} &=&\frac{1}{\sqrt{2}}\left( {\bf S}_{1}-{\bf S}
_{2}\right) =\frac{1}{\sqrt{2}}{\bf S}_{-},  
\label{new} \\
\sum_{n}in{\bf S}_{-n,n} &=&\sqrt{2}\left[ {\bf S}_{1}\times {\bf S}_{2}
\right] =\sqrt{2}{\bf T}.  
\nonumber
\end{eqnarray}
In terms of ${\bf\rm S}_{1,2}$, the effective Hamiltonian takes the form:
\begin{eqnarray}
&&H=\sum_{nks}\xi _{k}\psi _{nks}^{\dagger }\psi _{nks}
+K\left({\bf S}_{1}\cdot {\bf S}_{2}\right) 
-E_ZS^z_+
+\sum_{n}H_{n}, 
 \label{model}\\
&&H_{n} =J\left( {\bf s}_{nn}\cdot {\bf S}_{+}\right) +Vn\rho
_{nn}\left( {\bf S}_{1}\cdot {\bf S}_{2}\right)   
\label{Hn} \\
&&\quad\quad\quad
+\frac{I}{\sqrt{2}}\left[ \left( {\bf s}_{-n,n}\cdot {\bf S}_{-}\right)
+2in\left( {\bf s}_{-n,n}\cdot {\bf T}\right) \right].   
\nonumber
\end{eqnarray}
Here we introduced the particle and  spin densities in the continuum: 
\[
\rho _{nn}
=\sum_{kk's}\psi _{nks}^{\dagger }\psi _{nk's},\;
{\bf s}_{nn'}
=\sum_{kk'ss'}\psi_{nks}^{\dagger }
\frac{\bbox{\sigma }_{ss'}}{2}\psi _{n'k's'}.
\]
The bare values of the coupling constants are 
\begin{equation}
J=I=2V=2\left( t_L^2 + t_R^2\right)/E_C.
\label{equal}
\end{equation}
Note that the Schrieffer-Wolff transformation also produces a small 
correction to the energy gap $\Delta$ between the states $|1,1\rangle$ and 
$|0,0\rangle$, 
\begin{equation}
\Delta = E_{|1,1\rangle}-E_{|0,0\rangle} = K-E_Z,
\label{Delta}
\end{equation} 
so that $K$ differs from its bare value $K_0$, see (\ref{Hd}). However, 
this difference is not important, since it only affects the value of the control 
parameter at which the singlet-triplet transition occurs, but not the nature 
of the transition. 

We did not include into (\ref{model})-(\ref{Hn}) the free-electron 
Hamiltonian of the $\phi$-particles [see Eq.~(\ref{rotation})], as well as 
some other terms, that are irrelevant for the low energy renormalization. 
The contribution of these terms to the conductance is featureless at the 
energy scale of the order of $T_0$ (see the next section), 
where the Kondo resonance develops.  

At this point, it is necessary to discuss some approximations tacitly
made in the derivation of (\ref{model})-(\ref{Hn}).  First of all, we
entirely ignored the presence of many energy levels in the dot, and
took into account the low-energy multiplet (\ref{Basis}) only.  The
multi-level structure of the dot is important at the energies above
$\delta$, while the Kondo effect physics emerges at the energy scale
well below the single-particle level spacing \cite{multilevel}. The
high-energy states result merely in a renormalization of the
parameters of the effective low-energy Hamiltonian. One only needs to
consider this renormalization for deriving the relation between the
parameters $t_L$ and $t_R$ of the low-energy Hamiltonian (\ref{Hdot}),
(\ref{leads}) and (\ref{HT}) and the ``bare'' constants of the model
defined in a wide bandwidth $\epsilon_F$. On the other hand, using
the effective low-energy Hamiltonian, one can calculate, in principle,
the observable quantities such as conductance $G(T)$ and other
susceptibilities of the system at low temperatures ($T\ll\delta$), and
establish the relations between them, which is our main goal.

Note that the Hamiltonian (\ref{model})-(\ref{Hn}) resembles that of
the two-impurity Kondo model, for which 
$H_{n}=J_{n}\left(
{\bf s}_{nn}\cdot {\bf S}_{+}\right) 
+ I \left( {\bf s}_{-n,n}\cdot
{\bf S}_{-}\right) $ 
and the parameter $K$ characterizes the strength of the RKKY interaction 
\cite{ALJ}. It is known that the two-impurity Kondo model may undergo a 
phase transition at some special value of $K$ \cite{ALJ}. At this point, the 
system may exhibit non-Fermi liquid properties.  However, one can show 
\cite{PG}, using general arguments put forward in \cite{ALJ}, that the 
model (\ref{model})-(\ref{Hn}) does not have the symmetry that warrants 
the existence of the non Fermi liquid state. This allows one to apply the
local Fermi liquid description \cite{N} to study the properties of the
system at $T=0$. In the next section, we will concentrate on the 
experimentally relevant perturbative regime.

\subsection{Scaling Analysis}
\begin{figure}
\centerline{\epsfxsize=5.0cm
\epsfbox{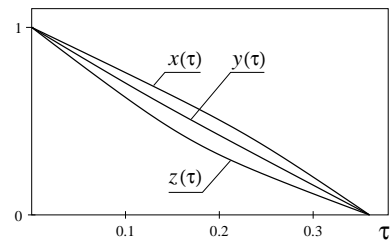}\vspace{1.5mm}}
\caption{ 
Numerical solution of the scaling equations. The RG equations
(\ref{Scaling}) are rewritten in terms of the new variable $\tau =\nu
J(0)\ln ({\delta}/D)$ and the new functions $x(\tau )=J(0)/J(\tau )$,
$y(\tau )=I(0)/I(\tau )$, $z(\tau )=V(0)/V(\tau )$ as
$dx/d\tau=-(1+x^2/y^2)$, $dy/d\tau=-(2y/x+y/z)$, $dz/d\tau=-4z^2/y^2$.
The three functions reach zero simultaneously at $\tau=\tau_{0}=0.36$.
}
\label{FigScaling}
\end{figure}

To calculate the differential conductance in the leading
logarithmic approximation, we apply the ``poor man's'' scaling
technique \cite{PWA}. The procedure consists of a perturbative 
elimination of the high-energy degrees of freedom and yields the 
and yields the set of scaling equations 
\begin{eqnarray}
&& dJ/d{\cal L} = \nu \left( J^{2}+I^{2}\right) ,
\nonumber \\  
&& dI/d{\cal L} = 2\nu I\left( J+V\right) ,
\label{Scaling} \\  
&& dV/d{\cal L} = 2\nu I^{2}
\nonumber  
\end{eqnarray}
for the renormalization of the coupling constants with the decrease of
the high energy cutoff $D$. Here ${\cal L}=\ln (\delta/D)$, and $\nu$ 
is the density of states in the leads; the initial value of $D$ is $D=\delta$, 
see the discussion after Eq.~(\ref{Delta}). The initial conditions for 
(\ref{Scaling}), $J(0)$, $I(0)$, and $V(0)$ are given by Eq.~(\ref{equal}).
The scaling procedure also generates non-logarithmic corrections to 
$K$. In the following we absorb these corrections in the re-defined value of 
$K$. Equations (\ref{Scaling}) are valid in the perturbative regime and as 
long as
\[
D \gg \left| K\right| , E_Z, T.
\] 
At certain value of ${\cal L}$, ${\cal L}= {\cal L}_{0}=\ln (\delta/T_{0})$, 
the inverse coupling constants simultaneously reach zero:
\[
1/J\left( {\cal L}_{0}\right) 
=1/I\left( {\cal L}_{0}\right) 
=1/V\left({\cal L}_{0}\right) 
=0. 
\]
This defines the
characteristic energy scale of the problem:
\begin{equation}
T_{0}=\delta\exp \left[ -\tau _{0}/\nu J\right]. 
\label{T0}
\end{equation}
Here $\tau _{0}$ is a parameter that depends
on the initial conditions and should be found numerically.  We
obtained $\tau _{0}=0.36$ (see Fig.~\ref{FigScaling}).  

It is instructive to compare $T_0$ with the Kondo temperature 
$T_K^{\rm odd}$ in the adjacent Coulomb blockade valleys with $N=odd$. 
In this case, only electrons from one of the two orbitals $n=\pm 1$ are involved 
in the effective Hamiltonian, which takes the form of the 1-channel $S=1/2$ 
Kondo model with the exchange amplitude 
$J_{\rm odd} = 4(t_L^2 + t_R^2)/E_C = 2J$, see Eq.~(\ref{equal}). Therefore,
$T_K^{\rm odd}$  is given by the same expression (\ref{T0}) as $T_0$, but with 
$\tau_0 = 1/2$. 
For realistic values of the parameters $T_0 = 300~mK$, $\delta = 3~meV$ 
we obtain $T_K^{\rm odd} \approx 120~mK$. This estimate is in a reasonable 
agreement with the experimental data, see Section \ref{DOTS} below.

The solution of the RG equations (\ref{Scaling}) can now be expanded
near ${\cal L}={\cal L}_{0}$. To the first order in 
${\cal L}_{0}-{\cal L}=\ln D/T_{0}$, we obtain
\begin{equation}
\frac{1}{\nu J({\cal L})}
= \frac{\sqrt{\lambda }}{\nu I({\cal L})}
= \frac{\lambda -1}{2\nu V({\cal L})}
= \left( \lambda +1\right) \ln (D/T_{0}),  
\label{ScalingSolution}
\end{equation}
where 
\[
\lambda =2+\sqrt{5}\approx 4.2.
\]
It should be emphasized that, unlike $\tau_0$, the constant $\lambda$
is universal in the sense that its value is not affected if the
restriction (\ref{tunneling}) is lifted \cite{future}.

Eq.~(\ref{ScalingSolution}) can be used to calculate the
differential conductance at high temperature 
$T\gg \left| K\right|, E_Z, T_{0}$.  In this regime, the coupling constants
are still small, and the conductance is obtained by applying a perturbation 
theory to the Hamiltonian (\ref{model})-(\ref{Hn}) with renormalized 
parameters (\ref{ScalingSolution}), taken at $D=T$, and using (\ref{formula}). 
This yields
\begin{equation}
G/G_0=\frac{A}{\left[\ln (T/T_{0})\right]^2},
\label{transition}
\end{equation}
where 
\[
A=\left( 3\pi ^{2}/8\right) \left( \lambda +1\right) ^{-2}
\left[1+\lambda +\left( \lambda -1\right) ^{2}/8\right] 
\approx 0.9
\]
is a numerical constant, and
\begin{equation}
G_{0}=\frac{4e^2}{h}
\left ( \frac{2 t_L t_R}{t^2_L + t^2_R} \right )^2 .
\label{G0}
\end{equation}

As temperature is lowered, the scaling trajectory (\ref{Scaling}) 
terminates either  at $D\sim {\rm max}\{|K|,E_Z\} \gg T_0$, or 
when the system approaches the strong coupling regime 
$D\sim T_0 \gg |K|,E_Z$. It turns out that the two limits of the theory, 
$E_Z \ll T_0$ and $E_Z \gg T_0$, describe two distinct physical situations,
which we will discuss separately.

\subsection{Singlet-triplet transition \label{TRANSITION}}
In this section, we assume that the Zeeman energy is negligibly small
compared to all other energy scales. At high temperature $T\gg |K|,T_0$,
the conductance is given by Eq.~(\ref{transition}).  At low
temperature $T\lesssim |K|$ and away from the singlet-triplet degeneracy
point, $|K|\gg T_0$, the RG flow yielding Eq.~(\ref{transition})
terminates at energy $D\sim |K|$. On the {\it triplet} side of the
transition ($K\ll -T_0$), the two spins ${\bf S} _{1,2}$ are locked 
into a triplet state. The system is described by the effective 2-channel
Kondo model with $S=1$ impurity, obtained from
Eqs.~(\ref{model})-(\ref{Hn}) by projecting out the singlet state and
dropping the no longer relevant potential scattering term:
\begin{equation}
H_{\rm triplet}=\sum_{nks}\xi _{k}\psi _{nks}^{\dagger }\psi_{nks}
+ J\sum_{n}\left( {\bf s}_{nn}\cdot {\bf S}\right);
\label{Htriplet}
\end{equation}
here $J$ is given by the solution $J({\cal L})$ of Eq.~(\ref{Scaling}), 
taken at ${\cal L} = {\cal L}^* = \ln(\delta/|K|)$, which corresponds 
to $D = |K|$. 

As $D$ is lowered below $|K|$, the renormalization of the
exchange amplitude $J$ is governed by the standard RG equation
\cite{PWA}
\begin{equation}
dJ/d{\cal L} = \nu J^{2},
\label{Tscaling} 
\end{equation}
where
${\cal L} = \ln (\delta/D) > {\cal L}^*$.
Eq.~(\ref{Tscaling}) is easily integrated with the result
\[
1/\nu J({\cal L}) - 1/\nu J({\cal L}^*)  = {\cal L} - {\cal L}^*.
\]
This can be also expressed in terms of the running bandwidth $D$ and the
Kondo temperature 
\[
T_k = |K|\exp\left[ -1/\nu J({\cal L}^*) \right]
\] 
as $1/\nu J({\cal L}) =\ln (D/T_{k})$.

Obviously, $T_k$ depends on $|K|$. Using asymptotes of $J({\cal L})$, 
see Eq.~(\ref{ScalingSolution}), we obtain the scaling relation
\begin{equation}
T_k/T_0 
= \left( T_0/|K|\right) ^{\lambda}.
\label{Tk}
\end{equation}
Eq.~(\ref{Tk}) is valid not too far from the transition point, where
the inequality
\begin{equation}
1\lesssim  |K|/T_0 \ll (\delta/T_0)^\mu, 
\; 
\mu \approx 0.24
\label{ineq}
\end{equation}
is satisfied. Here, $\mu$ is a numerical constant, which depends on 
$\tau_0$, and therefore is not universal [see the remark after 
Eq.~(\ref{ScalingSolution})]. For larger values of $|K|$ (but still smaller 
than $\delta$), $T_k \propto 1/|K|$ \cite{future}. Finally, for 
$|K| = \delta$, $T_k$ is given by Eq.~(\ref{T0}) with $\tau _{0} = 1$. 
According to (\ref{Tk}), $T_k$ decreases very rapidly with $|K|$.   
For example, for $T_0 = 300~mK$ and $\delta = 3~meV$ Eq.~(\ref{Tk}) 
describes fall of $T_k$ by an order of magnitude within the limits of its 
validity (\ref{ineq}). For $|K| =\delta $ one obtains $T_k \approx 5~mK$,
which is well beyond the reach of the present day experiments.  

For a given $|K|$, $T_0 \lesssim |K|\lesssim \delta$, the differential
conductance can be cast into the scaling form,
\begin{equation}
G/G_{0}=F\left( {T}/{T_{k}}\right) 
\label{triplet}
\end{equation}
where $F\left( {x}\right) $ is a smooth function that interpolates
between $F\left( 0\right) =1$ and 
$F\left( x\gg 1\right) =\left( \pi ^{2}/2\right) (\ln x) ^{-2}$.
It coincides with the scaled resistivity
\[
F(T/T_K)=\rho (T/T_K)/\rho(0)
\] 
for the symmetric two--channel $S=1$ Kondo model.  The conductance at
$T = 0$ (the unitary limit value), $G_{0}$, is given above in
Eq.~(\ref{G0}).

\begin{figure}
\centerline{\epsfxsize=6.0cm
\epsfbox{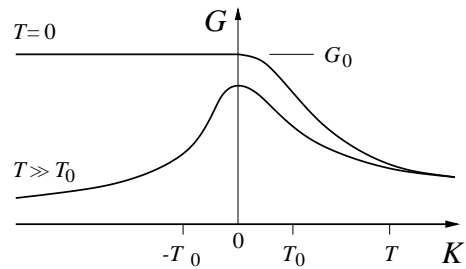}\vspace{1.5mm}}  
\caption{
Linear conductance near a singlet-triplet transition.  At high
temperature $G$ exhibits a peak near the transition point. At low
temperature $G$ reaches the unitary limit at the triplet side of the
transition, and decreases monotonously at the singlet side. The two
asymptotes merge at $K\gg T,T_0$. 
}
\label{overall}
\end{figure}

On the {\it singlet} side of the transition, $K\gg T_0$, the scaling
terminates at $D\sim K$, and the low-energy effective Hamiltonian is
\[
H_{\rm singlet}=\sum_{nks}\xi _{k}\psi _{nks}^{\dagger }\psi
_{nks} -\frac{3}{4}V \sum_{n}n\rho _{nn}, 
\]
where $V$ is $V({\cal L})$ [see Eq.~(\ref{Scaling})] taken at 
${\cal L} = {\cal L}^*$. The temperature dependence of the conductance 
saturates at $T\ll K$, reaching the value
\begin{equation}
G/G_0=\frac {B}{\left[\ln (K/T_0)\right] ^2},
\quad 
B=\left( \frac{3\pi }{8}\frac{\lambda -1}{\lambda +1}\right)^2
\approx 0.5 .
\label{singlet}
\end{equation}

Note that at $T=0$ Eqs.~(\ref{triplet}) and (\ref{singlet}) predict
different dependence on the parameter $K$ which is used for tuning
thorugh the transition. At positive $K$, conductance decreases with
the increase of $K$; at $K\ll -T_0$ conductance $G = G_0 $ and 
does not depend of $K$. Although there is no reason for the function 
$G(K)$ to be discontinious \cite{PG}, it is obviously a  
{\it non-analytical} function of $K$, see Fig.~\ref{overall}.  

The above results are for the linear conductance $G$. At $T = 0$, $G$ is 
a monotonous function of $K$, at high temperature $T \gg T_0$ the 
conductance develops a peak at the singlet-triplet transition point $K= 0$.
We now discuss shortly out-of-equilibrium properties. When the 
system is tuned to the transition point $K = 0$, the differential conductance 
$dI/dV$ exhibits a peak at zero bias, whose width is of the order of $T_0$. 
For finite $K$ the peak splits in two, located at {\it finite} bias $eV = \pm K$. 
The mechanism of this effect is completely analogous to the Zeeman 
splitting of the usual Kondo resonance \cite{Appelbaum},\cite{reviews}:  
$|K|$ is the energy cost of the processes involving a singlet-triplet transition. 
This cost can be covered by applying a finite voltage $eV = \pm K$, so that 
the tunneling electron has just the right amount of extra energy to activate 
the singlet-triplet transition prosesses described by the last two terms in 
(\ref{Hn}). The split peaks gradually disappear at large $|K|$ due to the 
nonequilibrium-induced decoherence \cite{MWL},\cite{lifetime}.

\subsection{Transition Driven by Zeeman Splitting \label{ZEEMAN}}
\begin{figure}
\centerline{\epsfxsize=4.2cm
\epsfbox{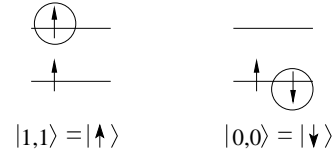}\vspace{2mm}}
\caption{ 
The ground state doublet in case of a large Zeeman splitting. The
states $|1,1\rangle$ and $|0,0\rangle$ differ by flipping a spin of a
single electron (marked by circles).}
\label{states}
\end{figure}

If the Zeeman energy is large, the RG flow (\ref{Scaling}) terminates
at $D\sim E_Z$. The effective Hamiltonian, valid at the energies
$D\lesssim E_Z$ is obtained by projecting (\ref{model})-(\ref{Hn})
onto the states $|1,1\rangle$ and $|0,0\rangle$. These states differ
by a flip of a spin of a single electron (see Fig.~\ref{states}), and
are the counterparts of the spin-up and spin-down states of $S=1/2$
impurity in the conventional Kondo problem. It is therefore convenient
to switch to the notations
\begin{equation}
|1,1\rangle = | \uparrow \rangle, 
\quad
|0,0\rangle = | \downarrow \rangle, 
\label{spins}
\end{equation}
and to describe the transitions between the two states in terms of the
spin-like operator 
\[
\widetilde{\bf S} = \frac{1}{2} \sum_{ss'} 
|s\rangle \bbox{\sigma}_{ss'}\langle s'| ,
\]
built from the states (\ref{spins}).

Projecting onto the sates (\ref{spins}), we obtain from (\ref{model})-(\ref{Hn})
\begin{eqnarray}
H=&&\sum_{nks}\xi _{k}\psi _{nks}^{\dagger }\psi _{nks}
+ \Delta \widetilde{S}^z 
\nonumber\\
&& +\sum_{n} \left [ 
Js^z_{nn} \left(\widetilde{S}^z +1/2\right)
+ Vn\rho_{nn} \left(\widetilde{S}^z -1/4\right)
\right ]
\label{projected} \\
&& -I\left( s^+_{1,-1}\widetilde{S}^-   +  {\rm H.c.}\right),
\nonumber
\end{eqnarray}
where $\Delta$ was introduced above in Eq.~(\ref{Delta}). It is now convenient 
to transform (\ref{projected}) to a form which is diagonal in the orbital indexes 
$n$. This is achieved simply by relabeling the fields according to
\begin{eqnarray}
&&\psi_{+1,k,\uparrow}= a_{k,\uparrow},
\;
\psi_{-1,k, \downarrow}= -a_{k,\downarrow}, 
\label{fields} \\
&&\psi_{-1, k,\uparrow}= b_{k,\uparrow},
\;
\psi_{+1,k, \downarrow}= -b_{k, \downarrow},
\nonumber
\end{eqnarray}
which yields
\begin{eqnarray}
H=&& H_0 + \Delta \widetilde{S}^z 
\nonumber \\
&&+ V_a s^z_a
 + J_z s^z_a \widetilde{S}^z  
+ \frac{1}{2} J_\perp \left ( s^+_a \widetilde{S}^- + s^-_a \widetilde{S}^+\right ) 
\label{kkondo} \\
&& +  V_bs^z_b +   J'_z s^z_b \widetilde{S}^z ,
\nonumber
\end{eqnarray}
where $H_0$ is a free-particle Hamiltonian for $a,b$ electrons, and  ${\bf s}_a$ 
is the spin density for $a$ electrons, 
${\bf s}_a=\sum_{kk'ss'}a_{ks}^{\dagger }
\left(\bbox{\sigma }_{ss'}/2\right)a _{k's'}$ 
(with a similar definition for ${\bf s}_b$).  The coupling constants in
(\ref{kkondo}),
\begin{eqnarray}
&& V_a = (J-V)/2,\; J_z= J+2V, \; J_\perp = 2I, 
\label{amplitudes} \\
&& V_b = (J+V)/2, \; J'_z= J-2V
\nonumber 
\end{eqnarray}
are expressed through the solutions of the RG equations
(\ref{Scaling}) taken at ${\cal L} =  {\cal L}^{**} = \ln(\delta/E_Z)$. 

The operators in  $b$-dependent part of (\ref{kkondo}) are not
relevant for the low energy renormalization. At low enough temperature 
(satisfying the condition 
$\ln(T/T_Z) \ll (\nu J'_z)^{-1},\, (\nu V_b)^{-1}$, 
where $T_Z$ is the Kondo temperature), their contribution to the 
conductance becomes negligible compared to the contribution from 
the $a$-dependent terms. This allows us to drop the
$b$-dependent part of (\ref{kkondo}). Suppressing the (now
redundant) subscript of the operators $s_a^i$, we are left with the
Hamiltonian of a one-channel $S=1/2$ anisotropic Kondo model,
\begin{equation}
H=H_0 + \Delta \widetilde{S}^z 
+ V_a s^z
 + J_z s^z \widetilde{S}^z  
+ \frac{J_\perp}{2} ( s^+ \widetilde{S}^- + {\rm H.c.}).
\label{kondo} 
\end{equation}
Eq.~(\ref{kondo}) emerged as a limiting case of a more general
two-channel model (\ref{model})-(\ref{Hn}). It should be noticed,
however, that the same effective Hamiltonian (\ref{kondo}) appears
when one starts with the single-channel model from the very beginning
\cite{Zeeman}.

A finite magnetic field singles out the $z$-direction, so that the 
spin--rotational symmetry is absent in (\ref{model})-(\ref{Hn}). This property 
is preserved in (\ref{kondo}). Indeed, even for $J_z = J_\perp$, 
Eq.~(\ref{kondo}) contains term $V_{\psi}s^z$ which has the meaning 
of a magnetic field acting {\it locally} on the conduction electrons at
the impuirity site.  The main effect of this term is to produce a
correction to $\Delta$, through creating a non-zero expectation value
$\langle s^z \rangle $ \cite{SW}. This results in a correction to $\Delta$. 
Fortunately, this correction is not important, since it merely shifts the 
degeneracy point. In addition, this term leads to insignificant corrections
to the density of states \cite{Zeeman}. 

Let us now examine the relation between $J_z$ and $J_\perp$. 
It follows from Eqs.~(\ref{equal}) and (\ref{amplitudes}), that at 
$E_Z = \delta$ the exchange is isotropic:  $J_z = J_\perp$. Moreover,
it turns out that if $E_Z$ is so close to $\delta$, that Eqs.~(\ref{Scaling}) 
can be linearized near the weak coupling fixed point ${\cal L} = 0$, the 
corrections to $J_z,J_\perp$ are such that the isotropy of exchange is 
preserved:
\begin{equation}
J_z=J_\perp =2J(0)\left [1 + \frac{3}{4} \nu J(0) \ln (\delta/E_Z) \right ],
\label{iso}
\end{equation}
where $J(0)$ is given by (\ref{equal}). This expression is valid as
long as the logarithmic term in the r.h.s. is small: 
$\nu J(0) \ln (\delta/E_Z) \ll 1$. 
Using (\ref{iso}) and (\ref{T0}), one obtains the Kondo temperature $T_Z$, 
which for  $J_z=J_\perp $ is given by
\[
T_Z= E_Z \exp[-1/\nu J_z] 
= E_Z (\delta/E_Z)^{3/8} (T_0/\delta)^{1/2\tau_0}
\]
Note that for $E_Z = \delta$, $T_Z$ coincides with the Kondo temperature 
$T_K^{\rm odd}$ in the adjacent Coulomb blockade valleys with odd number 
of electrons \cite{Zeeman}, see the discussion after Eq.~(\ref{T0}) above.

Note that the anisotropy of the exchange merely affects the value of $T_Z$ 
(which can be written explicitely for arbitrary $J_z$ and $J_\perp$ \cite{TW}). 
In the universal regime (when $T$ approaches $T_Z$), the exchange can be 
considered isotropic. This is evident from the scaling equations 
\cite{PWA}
\begin{equation}
dJ_z/d{\cal L} = \nu J_\perp^2,
\;
dJ_\perp/d{\cal L} = \nu J_z J_\perp, 
\;
{\cal L} > {\cal L}^{**} 
\label{RGzeeman} 
\end{equation}
where ${\cal L} > {\cal L}^{**} = \ln (\delta/E_Z)$, whose solution 
approaches the line $J_z = J_\perp$ at large ${\cal L}$.

According to the discussion above, the term $V_{\psi}s^z$ in 
Eq.~(\ref{kondo}) can be neglected. As a results, (\ref{kondo}) acquires 
the form of the anisotropic Kondo model, with $\Delta$ playing the part of 
the Zeeman splitting of the impurity levels. This allows us to write down the 
expression for the linear conductance at once. Regardless the initial 
anisotropy of the exchange constants in Eq.~(\ref{kondo}), the 
conductance for $\Delta=0$ in the universal regime (when $T$ approaches 
$T_Z$ or lower) is given by
\begin{equation}
G=G_{0Z}  f\left( T/T_Z \right),
\label{Gzeeman}
\end{equation}
where $f(x)$ is a smooth function interpolating between $f(0)=1$ and 
$f(x\gg 1) = (3\pi^2/16)(\ln x)^{-2}$. Function $f\left(T/T_Z \right)$ coincides 
with the scaled resistivity for the one-channel $S=1/2$ Kondo model and its 
detailed shape is known from the numerical RG calculations \cite{CHZ}.  
The conductance at $T=0$,
\begin{equation}
G_{0Z}=\frac{2e^2}{h}
\left ( \frac{2 t_L t_R}{t^2_L + t^2_R} \right )^2 ,
\label{G0Z}
\end{equation}
is by a factor of 2 smaller than $G_0$ [see Eq.~(\ref{G0})]; $G_0$ includes
contributions from two channels and therefore is twice as large as the
single-channel result (\ref{G0Z}). 
At finite $\Delta \gg T_Z$, the scaling trajectory (\ref{RGzeeman}) terminates
at $D \sim \Delta$. As a result, at $T \lesssim \Delta$ the conductance is 
temperature-independent, and for $J_z = J_\perp$
\[
G = G_{0Z}f(\Delta/T_Z) 
= G_{0Z}\frac{3\pi^2/16}{\left[\ln(\Delta/T_Z)\right]^2}. 
\]
The effect of the de-tuning of the magnetic field from the degeneracy
point $\Delta \neq 0$ on the differential conductance away from 
equilibrium is similar to the effect the magnetic field has on
the usual Kondo resonance \cite{Appelbaum},\cite{reviews}. For
example, consider the case, relevant for the experiments on the carbon
nanotubes, see section \ref{nanotubes}, when the exchange energy 
$E_S$ [see Eq.~(\ref{Hdot})] is negligibly small.  When sweeping 
magnetic field from $B = - \infty $ to $B = + \infty $, the degeneracy 
between the singlet state of the dot and a component of the triplet is 
reached twice, at $B = B^*$ and $B = - B^*$, when $|E_Z|\approx \delta$. 
If the field is tuned to $B=\pm B^*$, then the differential conductance 
$dI/dV$ has a peak at zero bias. At a finite difference $|B|-|B^*|$ this 
peak splits in two located at $eV= \pm g\mu_B (|B|-|B^*|)$. 

\section{Experiments \label{EXP}}

\subsection{GaAs quantum dots \label{DOTS}}

Here, we discuss the case of a quantum dot with $N = even$ in a
situation where the last two electrons occupy either a spin singlet or
a spin triplet state. The transition between singlet and triplet state
is controlled with an external magnetic field. The range of the
magnetic field is small ($B \sim 0.2~T$, $g\mu_B B \sim 5~\mu V$) such
that the Zeeman energy can be neglected and that the triplet state is
fully degenerate\cite{sasaki}. The theory for this situation was
described in section \ref{TRANSITION}.

\begin{figure}
\centerline{\epsfxsize=7.8cm
\epsfbox{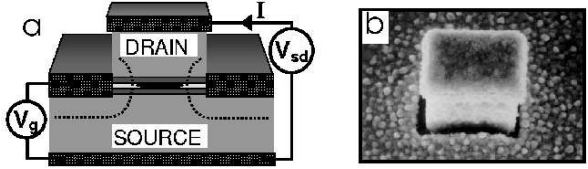}\vspace{2mm}}
\caption{ 
(a) Cross-section of rectangular quantum dot. The semiconductor material 
consists of an undoped AlGaAs(7nm)/InGaAs(12nm)/AlGaAs(7nm) double 
barrier structure sandwiched between n-doped GaAs source and drain 
electrodes. A gate electrode surrounds the pillar and is used to control the 
electrostatic confinement in the quantum dot. A dc bias voltage, $V$, is 
applied between source and drain and current, $I$, flows vertically through 
the pillar. The gate voltage, $V_g$, can change the number of confined 
electrons, $N$, one-by-one. A magnetic field, $B$, is applied along 
the vertical axis. \\
(b) Scanning electron micrograph of a quantum dot with dimensions 
$0.45\times 0.6 ~\mu m^2$ and height of $\sim0.5~\mu m$.
}
\label{dot1}
\end{figure}

\begin{figure}
\centerline{\epsfxsize = 4.5cm
\epsfbox{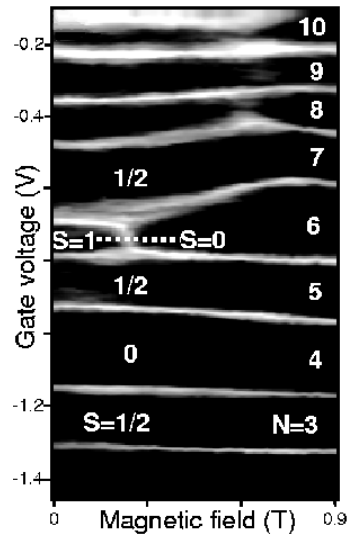}\vspace{3mm}}
\caption{ 
Gray-scale representation of the linear conductance $G$ versus the gate voltage 
$V_g$ and the magnetic field $B$. White stripes denote conductance peaks of 
height $\sim e^2/h$. Dark regions of low conductance indicate Coulomb blockade. 
The $N = 6$ ground state undergoes a triplet-to-singlet transition at 
$B=B^* \approx 0.22 ~T$, which results in a conductance anomaly inside the 
corresponding Coulomb gap.}
\label{dot2}
\end{figure}

The quantum dot has the external shape of a rectangular pillar 
(see Fig.~\ref{dot1}) and an internal confinement potential close 
to a two-dimensional ellipse \cite{Tarucha}. The tunnel barriers between the 
quantum dot and the source and drain electrodes are thinner than in other 
devices such that higher-order tunneling processes are enhanced. 
Fig.~\ref{dot2} shows the linear response conductance $G$ versus gate 
voltage $V_g$, and magnetic field $B$. Dark regions have low conductance 
and correspond to the regimes of Coulomb blockade for $N = 3$ to $10$. 
Light stripes represent Coulomb peaks as high as $\sim e^2/h$. The 
$B$-dependence of the first two lower stripes reflects the ground-state 
evolution for $N = 3$ and $4$. Their similar $B$-evolution indicates that the 
3rd and 4th electron occupy the same orbital state with opposite spin, which 
is observed also for $N = 1$ and $2$ (not shown). This is not the case for 
$N = 5$ and $6$. The $N = 5$ state has $S = 1/2$, and the corresponding 
stripe shows a smooth evolution with $B$. Instead, the stripe for $N = 6$ 
has a kink at $B=B^*\approx  0.22 ~T$. From earlier analyses \cite{Tarucha} 
and from measurements of the excitation spectrum at finite bias $V$ 
this kink is identified with a transition in the ground state 
from a spin-triplet to a spin-singlet. 

Strikingly, at the triplet-singlet transition 
(see Fig.~\ref{dot2}) we observe a strong enhancement of the conductance. 
In fact, over a narrow range around $0.22 ~T$, the Coulomb gap for 
$N = 6$ has disappeared completely. Note that the change in greyscale 
along the dashed line in Fig.~\ref{dot2} represents the variation of the 
conductance with the tuning parameter $K$, see Fig.~\ref{overall}.

\begin{figure}
\centerline{\epsfxsize=8.6cm
\epsfbox{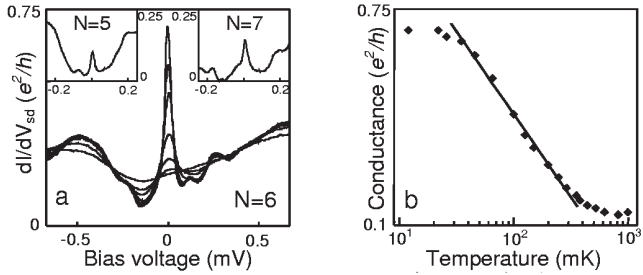}\vspace{3mm}}
\caption{(a) Kondo resonance at the singlet-triplet transition. 
The $dI/dV$ vs $V$ curves are 
taken at $V_g = -0.72~V$, $B = 0.21~T$ and for 
$T = 14, 65, 100, 200, 350, 520$, and $810~mK$. 
Kondo resonances for $N = 5$ (left inset) and 
$N = 7$ (right inset) are much weaker than for $N = 6$. \\
(b) Peak height of zero-bias Kondo resonance vs $T$ as obtained from (a). 
The line demonstrates a logarithmic $T$-dependence, which is 
characteristic for the Kondo effect. The saturation at low $T$ is likely 
due to electronic noise.
}
\label{dot3}
\end{figure}

To explore this conductance anomaly, Fig.~\ref{dot3}(a) shows the 
differential conductance, $dI/dV$ versus $V$, taken at $B$ and $V_g$ 
corresponding to the intersection of the dotted line and the bright 
stripe ($B=B^*$) in Fig.~\ref{dot2}. The height of the zero-bias resonance 
decreases logarithmically with $T$ [see Fig.~\ref{dot3}(b)]. These are 
typical fingerprints of the Kondo effect. From 
FWHM$\approx 30~\mu V \approx k_BT_0$, we estimate 
$T_0 \approx 350~mK$. Note that $k_BT_0/g\mu_B B^* \approx 6$ so 
that the triplet state is indeed three-fold degenerate on the energy scale 
of $T_0$; this justifies an assumption made in Section \ref{TRANSITION} 
above. Also note that some of the traces in Fig.~\ref{dot3}(a)
show small short-period modulations which disappear above 
$\sim 200~ mK$. These are due to a weak charging effect in the GaAs pillar 
above the dot \cite{B}. 

For $N = 6$ the anomalous $T$-dependence is found only when the 
singlet and triplet states are degenerate. Away from the degeneracy, the 
valley conductance increases with $T$ due to thermally activated transport. 
For $N = 5$ and $7$, zero-bias resonances are clearly observed [see insets 
to Fig.~\ref{dot3}(a)] which are related to the ordinary spin-$1/2$ Kondo 
effect. Their height, however, is much smaller than for the singlet-triplet Kondo 
effect.

\begin{figure}
\centerline{\epsfxsize = 4.2 cm
\epsfbox{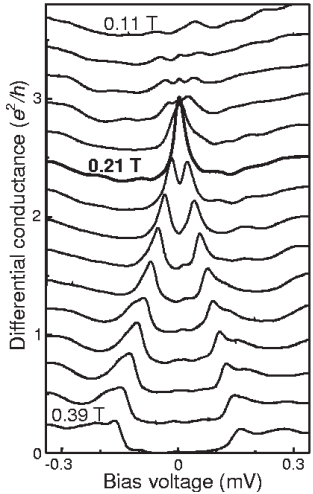}\vspace{3mm}}
\caption{ 
$dI/dV$ vs $V$ characteristics taken along 
the dotted line in Fig.~\ref{dot2} at equally spaced magnetic fields 
$B = 0.11,0.13,...,0.39~T$. Curves are offset by $0.25~e^2/h$. 
}
\label{dot4}
\end{figure}

We now investigate the effect of lifting the singlet-triplet degeneracy by 
changing $B$ at a fixed $V_g$ corresponding to the dotted line in 
Fig.~\ref{dot2}. Near the edges of this line, i.e. away from $B^*$, the 
Coulomb gap is well developed as denoted by the dark colours. The 
$dI/dV$ vs $V$ traces still exhibit anomalies, however, now at finite 
$V$ [see Fig.~\ref{dot4}]. 
For $B = 0.21~T$ we observe the singlet-triplet Kondo resonance at 
$V = 0$. At higher $B$ this resonance splits apart showing two peaks 
at finite $V$, in agreement with the discussion above (see Section 
\ref{TRANSITION}). For $B\approx 0.39~T$ the peaks have evolved 
into steps which may indicate that the spin-coherence 
associated with the Kondo effect has completely vanished. 
The upper traces in Fig.~\ref{dot4}, for $B < 0.21~ T$, also show 
peak structures, although less pronounced. \vspace{3.5mm}

\subsection{Carbon nanotubes \label{nanotubes}}
The situation in quantum dots formed in single-wall carbon nanotubes
\cite{tubes}, \cite{tans98},\cite{bockrath},\cite{cobden98} is rather
different from that in semiconductor quantum dots.  In nanotubes the
effect of magnetic field on orbital motion is very weak, because the
tube diameter $(\sim 1.4 ~ nm)$ is an order of magnitude smaller than
the magnetic length $l_B = (h/eB)^{1/2}\sim 10~nm$ at a typical
maximum laboratory field of $10~T$.  On the other hand, the $g$-factor
is close to its bare value of $g=2$, compared with $g = 0.44$ in GaAs.
Hence the magnetic response of a nanotube dot is determined mainly by
Zeeman shifts.  As a result, the spins of levels in nanotube dots are
easily measured \cite{tans98},\cite{cobden98}, and the ground state is
usually (though not always \cite{tans98}) found to alternate regularly
between an $S = 0$ singlet for even electron number $N$ and an $S
=1/2$ doublet for odd $N$ \cite{cobden98},\cite{david}.  Moreover,
singlet-triplet transitions in nanotubes are likely to be driven by
the Zeeman splitting rather than orbital shifts, corresponding to the
theory given in section \ref{ZEEMAN}.

\begin{figure}
\centerline{\epsfxsize=3.5cm
\epsfbox{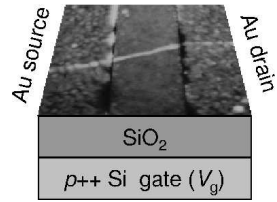}\vspace{1.5mm}}
\caption{ 
Schematic of a nanotube quantum dot, incorporating an atomic force 
microscope image of a typical device (not the same one measured here.)  
Bridging the contacts, whose separation is $200~ nm$, is a $2 ~nm$ thick 
bundle of single-walled nanotubes
}
\label{tube1}
\end{figure}

We discuss here the characteristics of a single-walled nanotube device with 
high contact transparencies, which were presented in more details in \cite{david}. 
The source and drain contacts are gold, evaporated on top of laser--ablation--grown 
nanotubes \cite{thess} deposited on silicon dioxide. The conducting silicon substrate 
acts as the gate, as illustrated in Fig.~\ref{tube1}. At room temperature the linear 
conductance $G$ is $1.6 ~e^2/h$, almost independent of gate voltage $V_g$, 
implying the conductance-dominating nanotube is metallic and defect free, and that 
the contact transmission coefficients are not much less than unity.  At liquid helium 
temperatures regular Coulomb blockade oscillations develop, implying the formation 
of a single quantum dot limiting the conductance.  However, the conductance in the 
Coulomb blockade valleys does not go to zero, consistent with high transmission 
coefficients and a strong coupling of electron states in the tube with the contacts. 

\begin{figure}
\centerline{\hspace{-3mm}\epsfxsize=7.8cm
\epsfbox{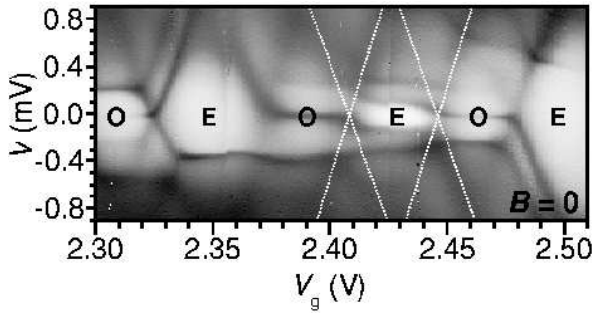}\vspace{2mm}}
\caption{ 
Grayscale plot of differential conductance $dI/dV$ (darker = more
positive) against bias $V$ and gate voltage $V_g$ at a series of
magnetic fields and base temperature ($\sim 75~ mK$).  Labels `E' and
`O' indicate whether the number of electrons $N$ in the dot is even or
odd (see text).  }
\label{tube2}
\end{figure}

Fig.~\ref{tube2} shows a grayscale plot of $dI/dV$ versus $V$ and $V_g$ 
over a small part of the full $V_g$ range at $B = 0$.  A regular series of faint 
"Coulomb diamonds" can be discerned, one of which is outlined by white 
dotted lines.  Each diamond is labeled either E or O according to whether $N$ 
is even or odd respectively, as determined from the effects of magnetic field.  
Superimposed on the diamonds are horizontal features which can be attributed 
to higher-order tunnelling processes that do not change the charge on the dot 
and therefore are not sensitive to $V_g$. 

\begin{figure}
\centerline{\epsfxsize=8.5cm
\epsfbox{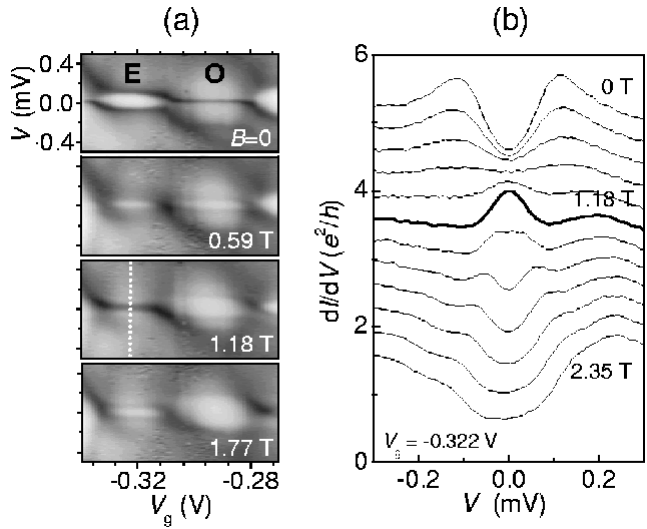}\vspace{2mm}}
\caption{ 
(a) Evolution with magnetic field of adjacent even (E) and odd (O)
features of the type seen in Fig.~\ref{tube2}.  \newline (b) $dI/dV$
vs $V$ traces at the center of the E region, at $V_g = - 0.322~ V$.
The trace at $B = B^* = 1.18~T$ (bold line) corresponds to the dotted
line in (a).  The traces are offset from each other by $0.4~ e^2/h$
for clarity.  }
\label{tube3}
\end{figure}

\begin{figure}
\centerline{\epsfxsize=8.5cm
\epsfbox{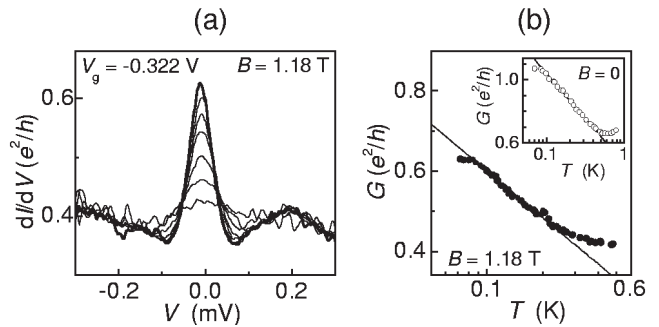}\vspace{2mm}}
\caption{ 
(a) Temperature dependence at $B = B^*$.  Here $T $= base (bold 
line), 100, 115, 130, 180, 230, and $350~mK$.  \newline
(b) Temperature dependence of the linear conductance $G$ ($dI/dV$ at $V=0$) at 
$B = B^* = 1.18 ~T$. For comparison, $G(T)$ in the center of one of the O-type ridges 
at $B = 0$ is shown in the inset.
}
\label{tube4}
\end{figure}

In Fig.~\ref{tube3}(a) we concentrate on an adjacent pair of E and O
diamonds in a magnetic field applied perpendicular to the tube.  At
$B=0$ the diamond marked with an `O' has narrow ridge of enhanced
$dI/dV$ spanning it at $V = 0$, while that marked with an `E' does
not. An appearence of a ridge at zero bias is consistent with
formation of a Kondo resonance which occurs when $N$ is odd (O) but
not when it is even (E).  This explanation is supported by the
logarithmic temperature dependence of the linear conductance in the
center of the ridge, as indicated in the inset to
Fig.~\ref{tube4}(b). At finite $B$, each zero-bias ridge splits into
features at approximately $V = \pm E_Z/e$ as expected for Kondo
resonances \cite{Appelbaum}.

On the other hand, in the E diamond the horizontal features appear at
a finite bias at $B=0$.  The origin of these features can be infered
from their evolution with a magnetic field: while the ridge in the O
region splits as $B$ increases, the edges of the E 'bubble' move
towards $V = 0$, finally merging into a single ridge at $B = B^* =
1.18 ~T$. Fig.~\ref{tube3}(b) shows the evolution with $B$ of the
$dI/dV$ vs $V$ traces from the center of the E region, and the
appearance of a zero-bias peak at around $B = 1.18~T$ (bold trace).
This matches what is expected for a Zeeman-driven singlet-triplet
transition in the $N=$ even dots (Section \ref{ZEEMAN}). Further
evidence that the peak is a Kondo resonance is provided by its
temperature dependence [Fig.~\ref{tube4}(a)], which shows an
approximately logarithmic decrease of the peak height (the linear
conductance) with $T$ shown in Fig.~\ref{tube4}(b).

Based on this interpretation we can deduce that for this particular
value of $N$ the energy gap separating the singlet ground state and
the lowest-energy triplet state is $\Delta_0 = g \mu_B B^* \approx
137~ \mu eV$.  At other even values of $N$ the lowest visible
excitations range in energy up to $\sim 400~\mu eV$.  For this device
$E_C \sim 500~\mu eV$. The energy gaps are therefore comparable with
the expected single-particle level spacing $\delta$, which is roughly
equal to $E_C/3$ in a nanotube dot\cite{bockrath}.

Note that the ridges at finite bias in Fig~\ref{tube3} in E valley are
more visible at $B=0$ than at $B=0.59~T$, halfway towards the
degeneracy point. A possible explanation is that at $B=0$ the triplet
is not split, and all its components should be taken into account when
calculating $dI/dV$ at $B=0$. This results in an enhancement of
$dI/dV$ at $B=0$, $eV=\delta$, as compared to the value expected from
the effective model of Section \ref{ZEEMAN}, which is valid in the
vicinity of $B=B^*$.

\section*{Conclusion}
Even a moderate magnetic field applied to a quantum dot or a segment
of a nanotube can force a transition from the zero-spin ground state
($S=0$) to a higher-spin state ($S=1$ in our case). Therefore, the
magnetic field may induce the Kondo effect in such a system. This is
in contrast with the intuition developed on the conventional Kondo
effect, which is destroyed by the applied magnetic field. In this
paper we have reviewed the experimental and theoretical aspects of the
recently studied magnetic--field--induced Kondo effect in quantum
dots. Clearly there is more territory to be explored in the remarkably
tuneable systems.

\section*{Acknowledgements}
We thank our collaborators from 
Ben Gurion University, Delft University of Technology, Niels Bohr Institute, 
NTT, and  University of Tokyo for their contributions. This work was 
supported by NSF under Grants DMR-9812340, DMR-9731756, by 
NEDO joint research program (NTDP-98), and by the EU via a TMR network.  
LG and MP are grateful to the Max Planck Institute for Physics of Complex 
Systems (Dresden, Germany), where a part of this paper was written, for the 
hospitality.

\end{multicols}


\begin{references}
\bibitem{B} The top contact is obtained by deposition of Au/Ge and annealing 
at $400~^{\circ}C$ for $30~s$. This thermal treatment is gentle enough to 
prevent the formation of defects near the dot, but does not allow the complete 
suppression of the native Schottky barrier. The residual barrier leads to 
electronic confinement and corresponding charging effects in the GaAs pillar.
\end{references}
\end{document}